\documentclass[aps,prl,twocolumn,groupedaddress]{revtex4}

\usepackage{graphicx}
\input{epsf}

\begin{document}

\title{Scanning Tunneling Microscopy and Spectroscopy study of charge inhomogeneities in bilayer Graphene}

\author{Shyam K. Choudhary and Anjan K. Gupta}
\affiliation{Department of Physics, Indian Institute of Technology Kanpur, Kanpur 208016, India.}
\date{\today}

\begin{abstract}
We report room temperature scanning tunneling microscopy and spectroscopy study of bilayer graphene prepared by mechanical exfoliation on SiO$_2$/Si surface and electrically contacted with gold pads using a mechanical mask. The bulk conductivity shows contribution from regions of varying electron density indicating significant charge inhomogeneity. Large scale topographic images show ripple like structures with a roughness of $\sim$1nm while the small scale atomic resolution images show graphite-like triangular lattice. Local dI/dV-V tunnel spectra have an asymmetric V-shape with the minima location showing significant spatial variation indicating inhomogeneity in electron density of order 10$^{11}$ cm$-2$. The minima in spectra at a fixed location also shifts linearly with the gate voltage with a slope consistent with the field induced carrier density.
\end{abstract}

\maketitle

Graphene has been a focus of research for past several years for its fascinating physics \cite{rmp-rev} and application potential arising from its unusual mechanical, optical, chemical, and electrical properties. The role of disorder and local electronic properties on bulk transport properties and the Dirac nature are some of the outstanding issues. The disorder is inevitable as one cannot have long range order in 2D and also due to other effects such as local Coulomb impurities in the supporting substrate and the electron-phonon interaction.

Scanning tunneling microscopy and spectroscopy (STM/S) is the most suitable tool for investigating the electronic nature of such defects as it can probe the local electronic density of states (DOS) by tunneling spectroscopy by measuring local dI/dV-V spectrum \cite{chen-book}. Recent attempts by some research groups on graphene have shown a variety of results. Crommie's group has been able to align the standard four probe Graphene Hall bar, made using the electron lithography techniques, with the STM tip. In these samples the atomic resolution images have been observed; however, the tunneling spectra do not agree with the Dirac behavior and show an energy gap of $\sim$100meV \cite{crommie-fet} at Fermi energy. STM/S work on thermally grown graphene and few layer graphene on SiC \cite{crommie-sic,sic1,sic2} also shows an energy gap of similar magnitude. Such graphene is known to be strongly coupled to the underlying SiC substrate. Another similar study by Deshpande et. al. does not show such a gapped behavior \cite{deshpande}. Andrei's \cite{andrei} group was able to find a small graphene on bulk graphite to show all the essential features of graphene including Dirac nature and Landau levels but without any control on electron density. Landau levels have also been seen on the bulk graphite surface \cite{LL-graphite}. On bilayer graphene, recent STM/S work by Deshpande et. al. \cite{deshpo-bilayer} shows charge impurities with no correlation with topography.
\begin{figure}
\includegraphics[width=3.3in]{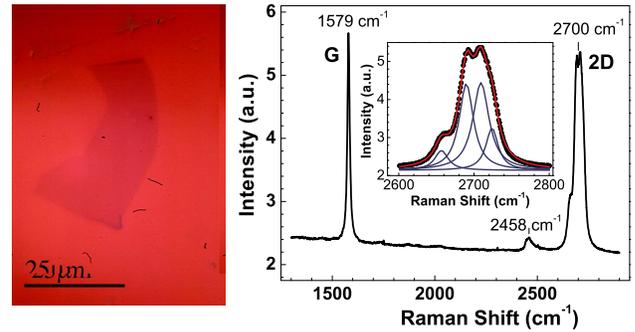}
\caption{Left picture shows the optical image of the graphene flake on SiO$_2$/Si identified as bilayer using the Raman Spectra, taken with a laser light of 514nm wavelength, as shown in the right plot. The 2D peak in this spectra near 2700cm$^{-1}$ can be fitted with four Lorenzians as shown in the inset ascertaining the flake as a bilayer graphene.}
\label{fig:optical-raman}
\end{figure}

In this paper, we present the STM study of bilayer graphene prepared by mechanical exfoliation on an insulating SiO$_2$ layer on a degenerately doped Si substrate acting as gate electrode. The small scale STM topographic images show atomically resolved surface while large scale images show ripple like structures. The local tunnel spectra (dI/dV-V) have an asymmetric parabolic shape with a finite zero bias conductance and with a minimum whose location along bias voltage axis varies spatially as well as with the gate voltage. The shift in minimum with the gate voltage is quantitatively consistent with the field induced carrier density but with an attenuation factor corresponding to a screening length of $\sim$1.2 nm. Both bulk conductivity and spatial variation in local tunnel spectra are compatible with charge inhomogeneities. The local conductance maps have a week correlation with the topography implying part of the topographic contrast is coming from charge inhomogeneities. We also find more charge inhomogeneity over longer length scales.

\begin{figure}
\includegraphics[width=3.3in]{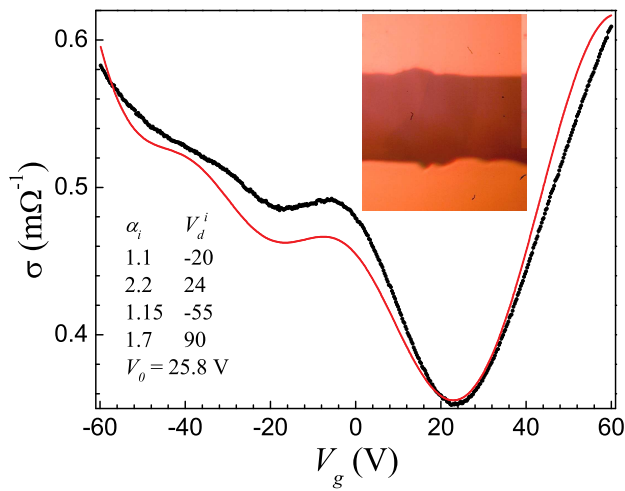}
\caption{Gate voltage dependent conductivity of bilayer graphene. The black line shows the experimental data while the red one is the model conductivity using Eq. \ref{eq:cond}, as described in the text, with the parameters listed in the plot.}
\label{fig:conductivity}
\end{figure}

The graphene flakes were prepared by mechanical exfoliation of graphite flakes \cite{graphite} on 300 nm thick SiO$_2$ on highly doped ($\rho <$0.005$\Omega$-cm) silicon substrate.  Bilayer graphene sheet, size ($\approx$ 42$\mu$m $\times$ 16$\mu$m) was optically identified \cite{optical-detection} as shown in Fig.\ref{fig:optical-raman}a. The central large portion of graphene was masked using a tungsten wire of 25$\mu$m diameter. A thin metal sheet with a slit of 1mm width and few mm length was also placed on top with slit length perpendicular to the wire to confine each contact to a narrow region. 100nm gold film followed by 10nm chromium was deposited after this masking to obtain two electrical contacts on either side of bilayer graphene as shown in Fig.\ref{fig:conductivity} inset. The graphene flake was confirmed to be a bilayer using Raman spectroscopy done in ambient conditions after the STM/S measurements. The Raman spectra, taken with 2.7mW Laser light of 514nm wavelength, are shown in Fig.\ref{fig:optical-raman}. We observed the G peak around 1580 cm$^{-1}$ and 2D band at 2700 cm$^{-1}$. Inset in Fig.\ref{fig:optical-raman} shows zoomed-in plot of 2D band with fitted to four Lorentzians at peak frequencies of 2657, 2689, 2709, and 2722 cm$^{-1}$. Four peaks in the 2D band of the Raman spectrum is the characteristic feature of bilayer graphene \cite{raman-rivew}. The absence of D peak at 1350 cm$^{-1}$ demonstrates the high quality of our samples.

All STM and transport data reported here were taken at room temperature and in 10$^{-4}$ mbar vacuum. This vacuum was obtained by cryo-pumping. Two probe transport measurements were done using a sinusoidal voltage of 597 Hz frequency and 100 mV amplitude across the graphene device in series with a 2.7 $\Omega$ resistance. The voltage across this resistance was measured using a lock-in amplifier for measuring the current. Gate voltage was applied to highly doped silicon electrode through a 270 k$\Omega$ series resistance. For STM measurements, one of the two electrodes (see inset of Fig.\ref{fig:conductivity}) of graphene device was connected to the STM bias circuitry, so that the tunneling current diffused in-plane through the gold film.  For aligning graphene with the STM tip, we have used a homemade 2D nano positioner \cite{nano-positioner} with a homemade STM with a design described elsewhere \cite{gupta-ng}. This STM uses a commercial electronics and control software \cite{rhk}. The electrodes defined in the previous step also served as guides for aligning the graphene with STM tip under an optical microscope. Electrochemically etched tungsten wire of 0.25 mm diameter were used as the STM tips. We also treat the etched tips with hydrofluoric acid to remove the oxide layer \cite{hf-cleaning}. The tunneling spectra (dI/dV-V) were acquired using the ac modulation technique with modulation amplitude of 20 mV. We kept the junction resistance value as 4G$\Omega$ all the local spectra taken at different gate voltage to keep the tip sample separation same for all the spectra and images. The simultaneous STM and STS images are obtained in constant current mode with the z-feedback time constant much larger than the time period of ac-modulation. In this mode opposite contrasts are observed in the simultaneous topographic and conductance images if the the surface is topographically flat but electronically inhomogeneous \cite{sts-contrast}. This also assumes the spectra to be monotonically rising with bias voltage. In our STM the bias voltage is applied to the sample while the tip is kept at (virtual) ground potential; thus the conductance at positive (negative) bias corresponds to electron DOS above (below) the Fermi energy.

\begin{figure}
\includegraphics[width=3.3in]{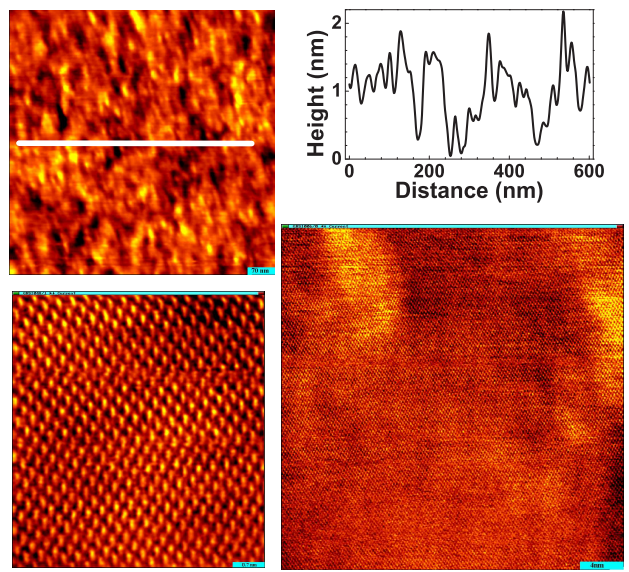}
\caption{Top left image shows a 690$\times$690nm$^2$ STM topograph of bilayer graphene taken at 0.4V bias and 0.1nA tunnel current with the topographic profile along the marked line shown in to top right plot showing the ripples. The bottom left image (5.4$\times$5.4nm$^2$) shows the atomic resolution image while the bottom right image (35$\times$35nm$^2$) shows the atomic resolution together with large scale ripples.}
\label{fig:topo}
\end{figure}

Fig.\ref{fig:conductivity} shows the variation of electrical conductivity of this bilayer graphene at room temperature with gate voltage. There is a prominent dip in the conductivity at positive gate voltage of 22.4 V showing the hole doped nature of this bilayer graphene. The conductivity value at this dip is 0.37m$\Omega^{-1}$, which is nearly twice of the minimum conductivity reported in graphene \cite{rmp-rev}, indicating good electrical contacts. Conductivity increases quadratically around this dip with gate voltage, until another smaller dip  appears at a gate voltage of about -20 V. This we believe is due to charge inhomogeneities in this large bilayer graphene flake. The  electron density, $n$ in the field effect geometry varies linearly with the gate voltage. Including the doping due to impurities or defects we can write,
\begin{eqnarray}
n=\frac{\kappa\epsilon_0}{ed}(V_g-V_d)
\label{eq:e-den}
\end{eqnarray}
Here $d$ (= 300nm) and $\kappa$ (= 4) are the thickness and dielectric constant of SiO$_2$, respectively, and $V_d$ is a voltage equivalent of the residual doping. $e$ is the magnitude of the electron's charge. For no doping, the density of states of bilayer graphene is quadratic \cite{bilayer-dos} with energy near E$_F$. Also there is a large DOS at E$_F$ so E$_F$ changes almost linearly with doping. Thus the DOS increases quadratically with doping giving a quadratic variation in conductivity with doping. Thus for fixed residual doping at a given location, the conductivity, $\sigma \propto [1+a(\frac{V_g-V_d}{V_0})^2]$. Assuming a simple configuration of N different regions, with different residual doping, connected in series between the two contacts, the expression for the average conductivity as a function of gate voltage is given by,
\begin{eqnarray}
\frac{1}{\sigma(V_g)}=\sum_{i=1}^{N} \frac{\alpha_i} {1+(V_g-V^i_d)^2/V_0^2}
\label{eq:cond}
\end{eqnarray}
Here $\alpha_i$ decides the relative fraction of the region with doping $V_i$. We can qualitatively produce the features of the measured conductivity by using four different doping regions with parameters as listed in Fig \ref{fig:conductivity}.  Although this simple model captures essential features of the experimental conductivity but it's an oversimplification and one needs to consider a resistor network where parallel regions also exist. More complications can arise if there is significant contribution from scattering at the boundary between regions. Although such charge inhomogeneities are confirmed by our STS spectra as discussed further, we cannot get the doping variation on the whole sample to calculate the detailed conductivity behavior.

\begin{figure}
\includegraphics[width=3.3in]{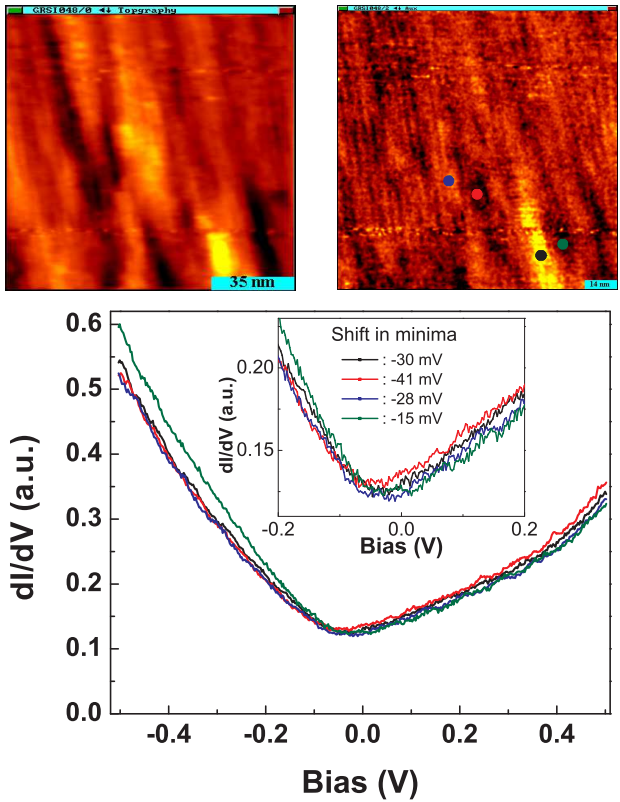}
\caption{The topographic (top left) and conductance (top right) images (121$\times$121nm$^2$) of bilayer graphene taken simultaneously in constant current mode at 0.4V bias and 0.1nA tunnel current.  The bottom plot shows the spectra taken at four different locations as marked in the conductance image. The inset in the lower plot shows the region close to minima to show the shift in the spectra. The lower spectra (with -0.05 offset) show a selection of spectra taken at various spots spanning several $\mu$m area (to be added).}
\label{fig:img-spect}
\end{figure}

Fig.\ref{fig:topo}a shows the topographic images of various sizes with the largest area (690$\times$690nm$^2$) image showing ripple like structures on a lateral length scale of $\sim$10-20nm and 1-2nm height. The smallest scale (5.4$\times$5.4nm$^2$) image shows atomically resolved surface while the intermediate size (35$\times$35nm$^2$) image shows simultaneously resolved ripples and the atomic features. Simultaneously taken topographic and conductance images at zero gate voltage are shown in Fig. \ref{fig:img-spect} together with some local spectra at the spots marked in the conductance image. We find that the spectra in the dark regions of the conductance image have a larger shift towards negative bias, i.e. E$_F$ is above the DOS minima or in other words it corresponds to more electron doping. In fact, most spectra at zero gate voltage show electron doped behavior with a shift as large as -100meV in some of the spectra taken at several $\mu m$ distance from this area. We estimate the spread in this shift to be $\sim$40meV from the average shift of $\sim$-40meV. This spread, from the discussion given later, is equivalent to the field effect doping by a gate voltage of $\sim$35V, which from Eq. \ref{eq:e-den} corresponds to an electron density variation of $\sim$10$^{11}$ cm$^{-2}$.

Other than this shift in minima the overall nature of the spectra remains same and thus we conclude that the conductance image contrast is predominantly arising from the variation in local electron density. Since these spectra are taken at room temperature, we do not see any gap due to large thermal smearing, which has been seen at low temperatures \cite{deshpo-bilayer}. This kind of electronic inhomogeneity can be attributed to either the charge fluctuations due to substrate \cite{ch-fluct} or the ripples \cite{deshpande}. We also find some correlation between the conductance and topographic images at small scales ($\sim$100nm) which makes us believe that some of the contrast in the topographic image is coming from this doping variation. We have found a cross-correlation coefficient in simultaneous STM and STS images as large as -0.2 and for a few scans we have also seen a positive cross-correlation coefficient. This indicates that there are generic topographic features, i.e. ripples, which could also be contributing to the electronic contrast.

\begin{figure}
\includegraphics[width=3.2in]{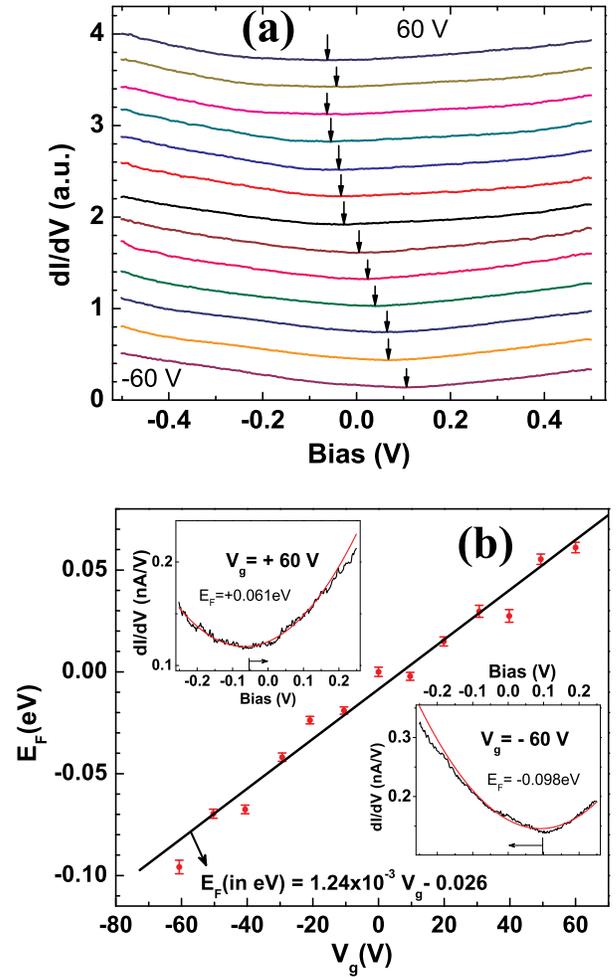}
\caption{(a) Spectra at different gate voltages (V$_g$) taken at the same position of an image (not shown). (b) Variation of E$_F$ as a function of the gate voltage. The two insets show a small portion of the spectra, taken at V$_g$=+60 an -60V, with the parabolic fitting to find the minima more accurately. The arrow at the bottom in the insets indicates the location of E$_F$ with respect to the spectral minima.}
\label{fig:Vg-dep}
\end{figure}

Fig. \ref{fig:Vg-dep}a shows the gate voltage dependence of the local dI/dV-V spectra with dI/dV being proportional to the DOS at energy eV with some thermal smearing. These spectra were acquired at the same point of an image (not shown). The spectra show the same shape except for a shift in the minima. The exact location of the minima was obtained by fitting a quadratic function in a small bias range ($\pm$100mV) about the minima as shown in Fig. \ref{fig:Vg-dep}b insets. This gives us the location of E$_F$ with respect to the minima as shown in Fig. \ref{fig:Vg-dep}b insets. The variation of the E$_F$ with gate voltage is shown in Fig.\ref{fig:Vg-dep}b; which shows a linear variation with gate voltage. The linear fitting gives a slope of (1.24$\pm0.1$)$\times 10^{-3}$.

To calculate this slope theoretically for a bilayer graphene with 2-fold spin and 2-fold valley degeneracy, the Fermi wave-vector is given by, $k_F = \sqrt{n\pi}$, with $n$ as the electron density. Assuming a tight binding parabolic band \cite{rmp-rev} near E$_F$, i.e., $\epsilon_{k \pm}= \hbar^2 v_F^2k^2/t_\bot$, we get $E_{F}= \pm\hbar^2 v_F^2n\pi/t_\bot$, with $t_\bot$ (= 0.4eV) as the inter-layer coupling parameter and $v_F$ (=1$\times$10$^{-6}$ m/s) as the Fermi velocity in single layer graphene. Using the field induced carrier density from Eq. \ref{eq:e-den} we get,
\begin{eqnarray}
E_{F}= \frac{\pi \kappa \epsilon_0\hbar^2 v_F^2} {e t_\bot d} (V_g-V_d)
\label{eq:EF-shift}
\end{eqnarray}
From this, $E_F$(in eV)=$\alpha(V_g-V_d)$ with $\alpha=\frac{\pi K \epsilon_0\hbar^2 v_F^2} {e^2 t_\bot d}$, which works out to be 2.53$\times$10$^{-3}$. The experimental value of the slope (see Fig. \ref{fig:Vg-dep}b) is smaller than this theoretical value by a factor of $\sim$0.5. This, we believe, is due to the screening effects, which will give rise to the field attenuation by a factor, $\beta=\exp({-t_{g}/\lambda})$ with $\lambda$ as the screening length and $t_g$ as the graphene thickness. Using $\lambda$ (=1.2 nm) \cite{screening} and $t_g$= 0.7 nm for bilayer graphene, we get $\beta$=0.56. Thus the slope should be $\beta\alpha$=1.41$\times$10$^{-3}$, which is in good agreement with the experimental value.

In conclusion, we have done a STM/S study with atomic resolution of gated bilayer graphene prepared by mechanical exfoliation. The STS spectra have an asymmetric V-shape with non-zero DOS near E$_F$ and with significant spatial variation in the location of minima. From this we find an inhomogeneity in electron density $\sim$10$^{11}$ cm$^{-2}$. The behavior of spectra with gate voltage is in quantitative agreement with the field induced doping together with the screening effects.
\section{Acknowledgements}
We thank Mandar Deshmukh for discussions and also for providing us with graphite crystals. We also thank Rajeev Gupta and Shyam Lal Gupta for Raman Spectroscopy. SKC acknowledges financial support from the UGC of the Government of India. Financial support from MHRD and DST of the Government of India is also thankfully acknowledged.

\end{document}